\documentclass[preprint, 12pt]{elsarticle}

\usepackage{amsmath,amsfonts,amsthm}
\usepackage{algorithmic}
\usepackage{algorithm}
\usepackage{array, longtable}
\usepackage[caption=false,font=normalsize,labelfont=sf,textfont=sf]{subfig}
\usepackage{textcomp}
\usepackage{stfloats}
\usepackage{url}
\usepackage{verbatim}
\usepackage{graphicx}
\usepackage{booktabs, arydshln}
\usepackage{float}
\usepackage{natbib}
\usepackage{color}
\usepackage{microtype}

\graphicspath{{images/}}

\begin{document}

\begin{frontmatter}

% ---------------- Title/Author ----------------
\title{VN-GraphGG: von Neumann--Mullins--based graph grain growth simulation}

\author[mymainaddress]{Yuntong Huang}

\author[secondaddress]{Zhihao Li}

\author[thirdaddress]{Shuyang Dai\corref{fn1}}
\ead{shuyang_dai@whu.edu.cn}

\author[mymainaddress]{Chang Ma\corref{fn1}}
\ead{changma@ust.hk}

\author[mymainaddress,fourthaddress,fifthaddress]{Yang Xiang\corref{fn1}}
\ead{maxiang@ust.hk}

\cortext[fn1]{Corresponding authors.}
\address[mymainaddress]{Department of Mathematics, The Hong Kong University of Science and Technology, Hong Kong, China}
\address[secondaddress]{Data Science and Analytics Thrust, Information Hub, The Hong Kong University of Science and Technology 
(Guangzhou), Guangzhou, China}
\address[thirdaddress]{School of Mathematics and Statistics, Wuhan University, Wuhan, Hubei, China}
\address[fourthaddress]{Department of Materials, The Hong Kong University of Science and Technology, Hong Kong, China}
\address[fifthaddress]{HKUST Shenzhen-Hong Kong Collaborative Innovation Research Institute, Futian, Shenzhen, China}

\begin{abstract}
Efficient simulations of grain growth are essential for investigating the process-structure-property relationships in polycrystalline
materials. We introduce a two-dimensional von Neumann-Mullins-based graph grain growth simulation (VN-GraphGG), which couples the 
grain-scale kinetics with microstructure reconstruction.
%separates the grain-area evolution from geometric reconstruction. 
The polycrystalline microstructure is represented as a graph with grain and junction nodes. The grain-area evolution 
on the grain graph is governed by the von Neumann--Mullins (VN) law, while a graph neural network is trained to reconstruct the microstructure from the graph. This separation allows kinetic coefficients to be varied without retraining the neural reconstruction model. Topological changes during grain growth such as disappearance of grains and changes in neighbor relations are handled 
straightforwardly through reconstruction and postprocessing. % without a separate topological solver.  
%Independent tests show accurate reconstruction and simulations support the recovery of the  prescribed kinetics and reproduce coarsening statistics. 
Simulations demonstrate accurate predictions of VN kinetics and coarsening statistics, and significant speedup over the phase-field simulations.
%Across five cases, the method reaches the half-grain-count threshold 
%with a mean wall-time speedup of $136.9\times$ over phase-field simulation. 
%Computational efficiency and the explicit kinetic control 
%enable ensemble studies of abnormal grain growth that examine how the magnitude and the distribution of kinetic heterogeneity influence microstructure evolution. 
The proposed method provides a foundation for efficient grain growth simulation based on grain-scale kinetics, opening 
the way to incorporate a broader range of physical mechanisms. An application example is presented for abnormal grain growth, in which kinetic heterogeneity significantly influences the grain microstructure evolution. 
\end{abstract}


\end{frontmatter}

% =========================================================
%
\section{Introduction}

Grain growth links thermal processing to the grain-size distribution, crystallographic texture and boundary network of
a polycrystal. These structural changes influence the strength and deformation mechanisms of nanocrystalline metals
\cite{Meyers2006}, the coercivity and permeability of nanocrystalline ferromagnets \cite{Herzer1990}. They also influence
the properties and reliability of polycrystalline thin films \cite{Thompson1990}, and microstructural development during 
ceramic and cermet processing \cite{FisherKang2019}. Grain growth follows recrystallization in many thermomechanical
treatments, connecting annealing conditions to the final grain structure and texture \cite{Doherty1997}. Efficient
simulation is therefore valuable for understanding both individual grain histories and the effects of initial
microstructure and kinetic parameters on grain-population statistics and topology \cite{Srolovitz1984Topology,LAZAR20116837,
Elsey2010}.

%{\color{red} In two dimensions, the von Neumann--Mullins (VN) law gives an explicit relation between the area rate of a grain and
%its number of sides \cite{von1952metal, mullins1956,palmer1994,MULLINS19986219}. For grain $g$ with area $A_g$, side number $n_g$, grain 
%boundary mobility $M_g$ and grain-boundary energy $\gamma$, the VN law is
%\begin{equation}
%    \frac{dA_g}{dt}=2\pi M_g\gamma\left(\frac{n_g}{6}-1\right)
%    =K_g(n_g-6),\qquad K_g=\frac{\pi M_g\gamma}{3}.
%    \label{eq:vn-law}
%\end{equation}
%The Mobility $M_g$ is commonly represented by the Arrhenius relation
%\[
%M_g(T)=M_{0,g}\exp[-Q_{M,g}/(RT)],
%\]
%whereas $\gamma$ may depend on temperature and boundary character \cite{burkeTurnbull1952,Rohrer2023}.
%This VN law and its three-dimensional generalization \cite{macphersonNeumannRelationGeneralized2007} 
%prescribe grain-scale kinetics, whereas they do not determine the geometry and topology of the evolving grains and the grain boundaries, such as the grain boundary profile, the grain disappearance or the changes 
%in neighbor relations. The grain growth simulation therefore requires reconstruction of the geometric and topological 
%evolution while preserving the prescribed kinetic response.}

Computational grain-growth methods have been developed along several complementary directions
\cite{FrostThompson1996,miodownik2002,chen2002,Rohrer2023}. Among the principal approaches, front-tracking and vertex
methods evolve boundary networks and their topological transitions explicitly
\cite{Kawasaki1989Vertex,CRPHYS_2010__11_3-4_265_0,LAZAR20116837}. Monte Carlo Potts models connect local lattice
updates to coarsening kinetics, grain-size distributions and topology
\cite{ANDERSON1984783,Srolovitz1984Topology}. Phase-field methods (PFM) represent grain boundaries as diffuse interfaces and have enabled
two- and three-dimensional studies of grain morphology and population statistics
\cite{FanChen1997,KrillChen2002,chen2002}. Level-set and diffusion-generated methods also provide front-capturing
strategies for complex or large grain networks \cite{Bernacki2011,Elsey2010}. Together, these methods have made
numerical simulation central to the study of grain-growth kinetics, statistics, and mechanisms. Their heavy computational cost, however, 
restricts the range of initial microstructure and kinetic conditions that can be explored across ensembles.

Machine-learning has been applied to both polycrystalline-property prediction and microstructure evolution
\cite{yang2021crnn, daiGraphNeuralNetworks2021, yanNovelPhysicsregularizedInterpretable2022,qin2023grainnn, fan2024gnn,
qin2024graingnn, TEP2025121486, peivaste2025fno, TIAN2026122153}. Graph neural networks (GNNs) update node representations
by aggregating information from neighboring nodes \cite{GNN2009}. This representation is well suited to polycrstals: grains 
can be represented as nodes, shared boundaries as edges, and geometric or physical features assigned to both. On this
basis, GNNs have been applied to provide interpretable predictions of polycrystalline properties \cite{daiGraphNeuralNetworks2021} 
and to accelerate grain-growth simulations through adaptive resolution \cite{fan2024gnn}. Based on dynamic graphs, GrainGNN \cite{qin2024graingnn} is developed to learn interface motion and topological changes for three-dimensional grain microstructure evolution, trained on phase field simulation data. A hybrid convolutional--graph neural network is proposed to learn grain-growth evolution in a
compressed representation from kinetic Monte Carlo data \cite{TIAN2026122153}. However, when learned evolution updates are applied iteratively,
the approximation errors can accumulate along the predicting trajectories. The applications of neural evolution surrogates are therefore limited.%Encoding the kinetics in the predictor also makes it challenging to impose and test a prescribed change in mobility independently.

A fundamental kinetic relation in two-dimensional grain growth is the von Neumann--Mullins (VN) law \cite{von1952metal, mullins1956,palmer1994,MULLINS19986219}, which gives an explicit relation between the area-change rate of a grain and
its number of sides. For a grain $g$ with area $A_g$, side number $n_g$, grain boundary mobility $m_g$, and grain-boundary 
energy $\gamma$, the VN law is
\begin{equation}
    \frac{dA_g}{dt}%=2\pi M_g\gamma\left(\frac{n_g}{6}-1\right)
    =M_g(n_g-6),
    \label{eq:vn-law}
\end{equation}
where $ M_g=\frac{\pi m_g\gamma}{3}$, and
the grain-boundary mobility $m_g$ 
%is commonly represented by the Arrhenius relation
%$M_g(T)=M_{0,g}\exp[-Q_{M,g}/(RT)$, 
%whereas $\gamma$ may depend on temperature and boundary character 
depends on temperature \cite{burkeTurnbull1952,Rohrer2023}. The VN law and its three-dimensional generalization \cite{macphersonNeumannRelationGeneralized2007} describe the grain-scale kinetics but do not specify
the evolving microstructure.
A Monte Carlo framework \cite{kim2023statistics} is established for evolving grain statistics, combining VN-law area evolution 
with a neural network for topological transformations. Such a statistical method does not supply a polycrystalline field
for spatial analysis.

%Coupling an explicit VN update to learned spatial reconstruction offers both independently specified kinetics and a resolved 
%microstructure while confining learning to geometric reconstruction.
We introduce VN-GraphGG that couples the VN law based evolution with grain microstructure reconstruction. Our method consists of the following three components for the evolution in each time step:
%{\color{red} 1. Incorporating physics via VN dynamics. 2. Reconstructing grain network instead of topological events.}
(i) A %labeled 
grain microstructure field is first converted into a grain graph with the information of grain-grain and grain-junction connectivity, grain areas, and other features. (ii) The VN law is then applied to evolve the grain areas. (iii) A neural graph-to-field reconstruction model is designed to recovers the grain microstructure for the 
next step evolution. A key feature of our method is the natural handling of topological events within this framework: the disappearance of grains 
and changes in neighbor relations during the grain microstructure evolution are taken care of by the reconstruction block with a post-processing treatment. %Graph re-extraction 
%incorporates these changes into the evolving grain network, and no separate topological solver is required. 
%{\color{red} Another advantage of our method lies in the effective 
%kinetic coefficient, which can be prescribed in the VN update, while the reconstruction model remains fixed after training.}
%Different initial fields and coefficients conditions can therefore be simulated without retraining the reconstruction model.
%Once trained, our model can predict grain growth from any initial grain microstructure and any kinetic coefficients of individual grains.
%Extending its labeled-field interface to segmented experimental microstructures provides a natural direction for future work.
The separation of kinetics and microstructure reconstruction enables the method to be applied generally to grain growth from any initial grain microstructure and any kinetic coefficients of individual grains, once the neural reconstruction model is trained. Since the VN law prescribes the grain area at each
step, there is no need to learn grain growth kinetics with the neural model. This helps reduce the accumulation of errors, compared to surrogates that learn the full evolution.

Simulations using our VN-GraphGG 
demonstrate accurate predictions of the VN kinetics and coarsening statistics of grain growth. (Recall that in our VN-GraphGG, an approximation is made that the VN law is applied to the reconstructed grain microstructure at each time step.)
 Especially, tests on the trained graph-to-field model show accurate reconstruction of grain geometry and adjacency. 
Comparisons with phase field simulations show that our VN-GraphGG gives similar bulk coarsening behaviors %at comparable grain counts 
and significant reduction of computational cost by two orders of magnitude.

Our method %combines the VN law with a learned spatial reconstruction, 
provides grain-scale kinetic controls and efficient simulations for grain growth. 
The separation of kinetics and microstructure reconstruction makes our method particularly suited to
controlled ensemble studies of kinetic heterogeneity effects on grain growth. 
%The model permits initial
%microstructures, mobility maps and kinetic hypotheses to be varied without retraining the reconstruction model, making
%population-level sensitivity and uncertainty accessible alongside individual grain histories. 
We demonstrate this capability by abnormal grain growth (AGG), in which initial grain sizes, pinning and heterogeneous grain 
boundary properties can significantly change the evolution of a small subset of grains
\cite{FisherKang2019, rollett1989,rios1997,razzak2012,decost2017}. Previous machine-learning studies treated AGG as a
rare-event prediction problem \cite{cohnGraphConvolutionalNetwork2024,zhouLearningPredictRare2025}. In our proposed method, we use ensemble
simulations to examine how controlled changes in kinetic coefficients influence the occurrence of AGG.
The framework is not limited to AGG; with appropriate kinetic rules, %attributes and validation,
the same neural network architecture can be generalized to study heat-treatment pathways, thin-film texture evolution, nanocrystalline stability
and other grain-boundary engineering tasks \cite{Thompson1990,Chookajorn2012,Darling2016,WatanabeTsurekawa1999}.

%{\color{red} The  efficiency, together with the explicit kinetic control, enables ensemble AGG studies across different high-coefficient fractions and coefficient contrasts.} The VN-GraphGG  method provides a basis for efficient grain growth simulation based on grain-scale kinetics and future investigation of process-structure-property relationships in polycrystalline materials.

%More broadly, the proposed model provides a flexible platform for future grain-growth
%simulations under diverse kinetic laws and microstructural fields.

% =========================================================
\section{Methods}

\begin{figure}[!t]
  \centering
  \includegraphics[width=\linewidth]{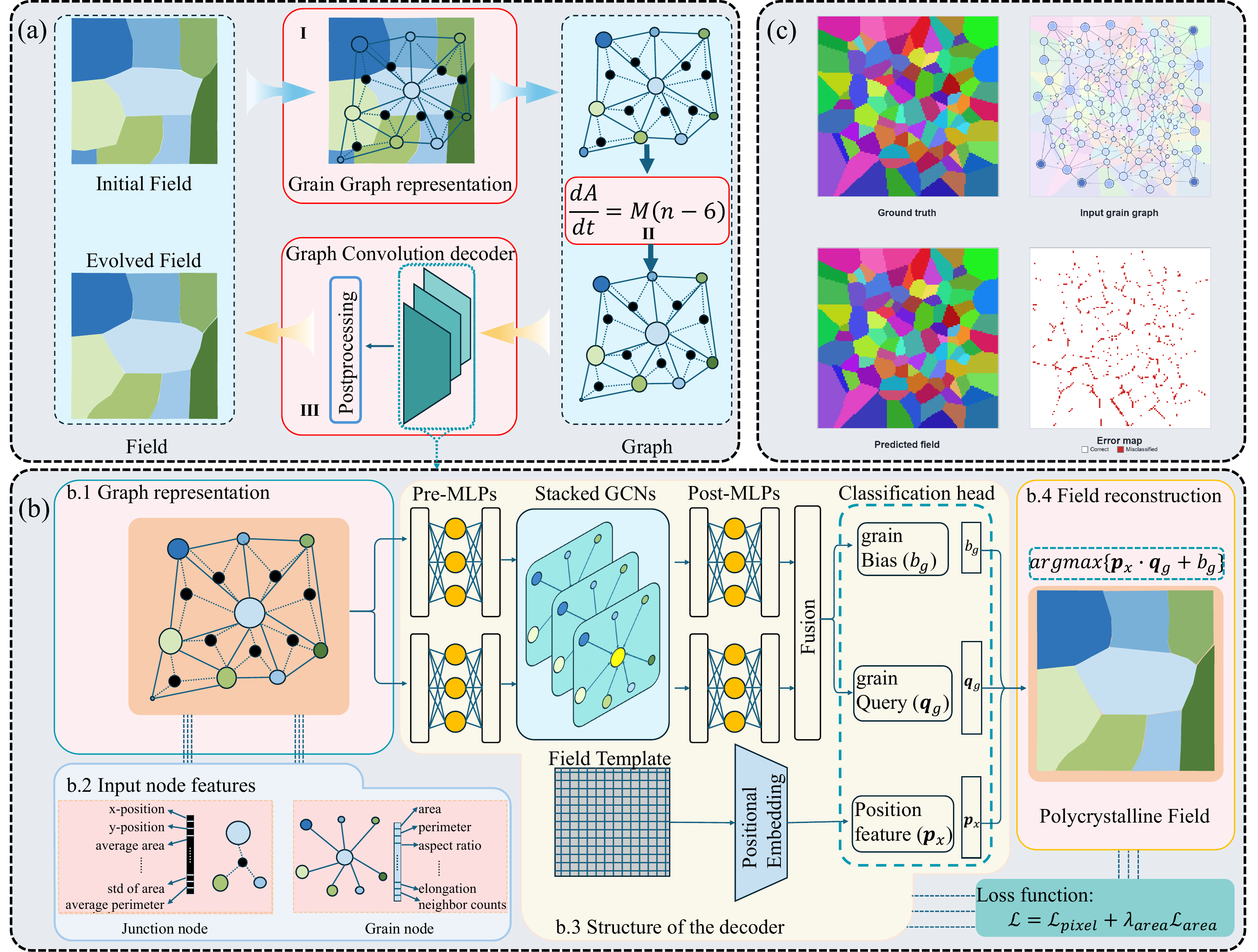}
  \caption{\textbf{The framework of VN-GraphGG.} (a) The proposed workflow for simulating grain growth: \uppercase\expandafter{\romannumeral1}, a 
  polycrystalline field is converted into a grain graph; \uppercase\expandafter{\romannumeral2}, grain areas are advanced by the prescribed VN law; and \uppercase\expandafter{\romannumeral3}, the learned reconstruction model and postprocessing supply the field for the next step.
  (b) The inputs (b.1 and b.2), structure (b.3), outputs (b.4), and loss functions of the reconstruction model. (c) A representative example 
  from the test set shows the ground truth field, corresponding graph, reconstructed field and pixel wise classification error.}
  \label{fig:framework}
\end{figure}

Our VN-GraphGG simulation method consists of three blocks: (I) Grain graph representation, (II) VN law evolution, and (III) %Graph convolution decoder. 
Grain field reconstruction.
In each time step, the grain microstructure passes through the three blocks consecutively, resulting in the grain microstructure at the next time step. See the overall workflow of VN-GraphGG in Fig.~\ref{fig:framework}a and the detailed architecture of the reconstruction model in Fig.~\ref{fig:framework}b.

%The proposed grain growth method couples the VN law based evolution with learned spatial reconstruction through an iterative framework (Fig.~\ref{fig:framework}a,b). Each iteration begins with a polycrystalline field, which is converted into a graph of grains and junction nodes. The VN law then is applied to evolve normalized grain areas on the grain nodes, and a learned graph-to-field model reconstructs the corresponding spatial field. Grain disappearance and changes in neighbor relations arise through reconstruction and postprocessing. Graph re-extraction then incorporate those events into the grain network. No separate topological solver is required. 

\subsection{Graph representation}

We start with a $N\times N$ microstructure field with $N^0$ grains in which each pixel $x$ stores an integer index of its grain,
$I(x)$, ranging from 1 to $N^0$. The field $L$ therefore is a $N\times N$ matrix consisting of $I(x)$, which is vectorized and converted into
an undirected graph $G$, containing grain nodes, junction nodes, grain-to-grain edges, and grain-to-junction edges
(Fig.~\ref{fig:framework} b.1 and b.2). We apply a numerical method to construct this grain graph \cite{daiGraphNeuralNetworks2021,
zhouLearningPredictRare2025}. In the present study, $N=128$. More details about construction algorithms, feature lists are included in Appendix A. 

Each distinct grain index defines one grain node. A 13-dimensional feature vector describes its geometry and local connectivity, 
including area, perimeter, and so on. These features are calculated from pixels carrying that index
and its connections. Grain-to-grain edges are determined by scanning adjacent pixels. When adjacent pixels have different indices, 
an undirected edge is created between the corresponding grain nodes.

Junction nodes are created at interior boundary pixels whose neighborhoods contain at least three grains \cite{Yang2024TripleJunction}. Each junction node has nine features, describing positions, connected-grain counts, and statistics
of connected-grains. The grain-to-junction edge connects the node to the 
grains within its neighborhood. 

In summary, the grain graph $G$ from a given field $L$ consists of three lists, which are the grain node list, the junction 
node list, and the edge list. In the grain node list, an integer array is used to record grain indices and the features are given as $\mathbb R^{N^0\times 13}$ matrices. The junction features are stored as $\mathbb R^{J\times 9}$ matrices, where $J$ is the number of junction nodes, for the junction node list. The edges are saved as pairs of node indices and grouped by relation, such as 
grain-to-grain edges and grain-to-junction edges. These nodes and edges encode grain geometry, grain adjacency, and junction 
information in polycrstalline structures. 

\subsection{Kinetic updating by VN law}

In block \uppercase\expandafter{\romannumeral2}, we apply the VN law to update the area feature on each grain node, then adjust the 
associated grain and junction features. In grain graphs, for a grain $g$, its normalized grain area $a_g=A_g/A_\Omega$ is stored as a feature, where $A_g$ is measured area in the field and $A_\Omega=N^2$ is the domain area. Dividing Eq.~\eqref{eq:vn-law} by 
$A_\Omega$ gives
\begin{equation}
    \frac{da_g}{dt}=M_{\mathrm{eff},g}(n_g-6),\qquad
    M_{\mathrm{eff},g}=\frac{M_g}{A_\Omega}
    =\frac{\pi m_g\gamma}{3A_\Omega},
    \label{eq:vn-normalized}
\end{equation}
where $t$ denotes time in the method and $M_{\mathrm{eff},g}$ is the normalized effective 
kinetic coefficient, with units of inverse time. 

Let $t_k=k\Delta t$ denote time at step $k$ and $\Delta t$ is the time-step size. The grain graph at this step is $G^k$, which
provides the current area and neighbor count for each grain.  $V_g^k$ contains distinct grain nodes
adjacent to the grain $g$. Its side number then is $n_g^k=|V_g^k|$. An explicit Euler 
discretization of Eq.~\eqref{eq:vn-normalized} defines the prescribed area
\begin{equation}
  \widetilde a_g^{k+1}=\max\left\{0,
  a_g^k+M_{\mathrm{eff},g}\Delta t(n_g^k-6)\right\},
  \label{eq:discrete-vn}
\end{equation}
where $a_g^{k}$ denotes the normalized area of the grain $g$ measured from the current microstructure field. The lower
bound of zero prevents negative prescribed areas.

We then update the perimeter of grains assuming that there is a similar scaling between the perimeter and area. Let 
$P_g^k$ denote the perimeter of the grain $g$ measured from the field and $p_g^k=P_g^k/(4\times N)$ is its normalized perimeter. 
We scale the normalized perimeter by the square root of the prescribed-to-current area ratio:
\begin{equation}
  \widetilde p_g^{k+1}=\max\left\{10^{-4},
  p_g^k\sqrt{\frac{\widetilde a_g^{k+1}}{\max(a_g^k,10^{-8})}}\right\}.
  \label{eq:perimeter-target}
\end{equation}
This scaling gives an estimate of the perimeter for reconstruction and preserves $p_g/\sqrt{a_g}$ when the numerical lower 
bounds are inactive. The area denominator and perimeter are regularized by the lower bounds $10^{-8}$ and $10^{-4}$, respectively. 

The area and perimeter features on each grain node now are replaced by the prescribed value $\widetilde a_g^{k+1}$ and $\widetilde p_g^{k+1}$. For each junction node, we recompute and update the related features, as the statistics of connected grains. Except
these, the graph connectivity, node positions and the remaining features are fixed at this block. The updated grain graph 
$\widetilde G^{k+1}$ is then passed into a decoder for reconstruction.

\subsection{Field reconstruction}

Block \uppercase\expandafter{\romannumeral3} consists of two components: a neural reconstruction model and a postprocessing step. The graph-to-field model is the first part and it maps a grain graph $G$ to its microstructure field $L$
(Fig.~\ref{fig:framework} b). We implement the decoder with heterogeneous graph-convolutional neural networks (GCNs) \cite{kipf2017iclr-semi} 
, which takes the graph $G$, including its connectivity and node features, as input 
(Fig.~\ref{fig:framework} b.3). The decoder handles the grain and junction nodes separately and processes different
edges using separate learned transformation \cite{heteroConv2018,morris2019graphconv}. 
The decoder outputs a $N\times N$ microstructure field
in which each pixel is classified to a grain index (Fig.~\ref{fig:framework} b.4).

Preprocessing multilayer perceptrons (Pre-MLPs) separately encode the grain features and junction features to latent 
vectors. Three graph-convolution layers with residual connections (Stacked GCNs) then are applied to 
exchange information between neighboring grains and between grains and their connected junction nodes. We use a
separate graph-convolution operator for each edge type \cite{morris2019graphconv} and sum these information as the
vectors of grain and junction nodes. Post-message-passing MLPs (Post-MLPs) further are applied to the grain 
and junction vectors separately. For each grain, the vectors of its connected junction nodes are averaged, which then 
is concatenated with the grain vector. The resulting vector is passed through the fusion layer and mapped into a grain 
query vector $\mathbf q_g$ and a scalar biase $b_g$ for the grain $g$. 

In parallel, the normalized coordinates of each pixel in a field template undergo Fourier positional encoding. A two-layer
two-layer MLP maps these encoded coordinates to a position vector $\mathbf p_x$ at each pixel $x$, with the same dimensional 
as the grain query vector $\mathbf q_g$. Following the query-based assignment strategy \cite{cheng2021maskformer}, the
decoder assigns grain index $I(x)$ to the 
each pixel $x$ with highest score $s_{xg}=\mathbf{p}_x^{\mathsf T}\mathbf q_g+b_g$ (Fig.~\ref{fig:framework} b.4),
\begin{equation}
    \qquad I(x)=\underset{g}{\operatorname{arg\,max}}\;s_{xg}=\underset{g}{\operatorname{arg\,max}}\;\mathbf p_x^{\mathsf T}\mathbf q_g+b_g,
    \label{eq:assignment}
\end{equation}
The resulting grain indices form the reconstructed microstructure field $L$. Appendix B specifies the architecture, 
dataset construction, training history, optimization settings and checkpoint-selection procedure.

During training, the scores $s_{xg}$ defined above are converted into grain-assignment probabilities using
\begin{equation}
P_{xg}=
\frac{\exp(s_{xg})}
{\sum_{h\in\mathcal G}\exp(s_{xh})},
\end{equation}
where $\Omega$ is the set of pixels and $\mathcal G$ is the set of grains in a field. Let $I_x\in\mathcal G$ denote the ground truth of pixel $x$. The pixel-classification loss is
\begin{equation}
\mathcal L_{\mathrm{pix}}=
-\frac{1}{|\Omega|}
\sum_{x\in\Omega}
\left(1-P_{x,I_x}\right)^\gamma
\log P_{x,I_x},
\end{equation}
where an exponent $\gamma=2$ is used.

To constrain the reconstructed grain sizes, we additionally define the reference and predicted soft area fractions as
\begin{equation}
a_g^{\mathrm{ref}}=
\frac{1}{|\Omega|}
\sum_{x\in\Omega}\mathbf 1\{I_x=g\},
\qquad
\widehat a_g^{\mathrm{soft}}=
\frac{1}{|\Omega|}
\sum_{x\in\Omega}P_{xg}.
\end{equation}
The relative soft grain-area loss is
\begin{equation}
\mathcal L_{\mathrm{area}}=
\frac{1}{|\mathcal G|}
\sum_{g\in\mathcal G}
\left[
\frac{\widehat a_g^{\mathrm{soft}}-a_g^{\mathrm{ref}}}
{\max\!\left(a_g^{\mathrm{ref}},10^{-6}\right)}
\right]^2.
\end{equation}
The decoder is trained using the combined objective,
\begin{equation}
\mathcal L=
\mathcal L_{\mathrm{pix}}
+\lambda_{\mathrm{area}}\mathcal L_{\mathrm{area}},
\qquad
\lambda_{\mathrm{area}}=10.
\end{equation}
The first term promotes accurate pixelwise assignments, whereas the relative area term gives grains 
comparable importance regardless of their size and encourages consistency between the ground truth and 
reconstructed grain areas. The resulting field $L$ is then passed to the postprocessing 
procedure described below.

The second part in block \uppercase\expandafter{\romannumeral3}, postprocessing, helps handle the topological changes, such as disappearances of grains and changes in neighbor relations. Grain occupying fewer than five pixels are removed, and 
their pixels are reassigned to neighboring grains. To reduce discrepancies 
between reconstructed grain area and the VN-prescribed one, boundary pixels are also reassigned from grains exceeding their VN-prescribed areas to their neighboring grains. Changes in grain adjacency can arise during 
this block \uppercase\expandafter{\romannumeral3} and are incorporated into the graph in block \uppercase\expandafter{\romannumeral1} at the next step.

Algorithm~\ref{alg:simulation-cycle} summarizes the workflow of simulation with VN-GraphGG. Here $L^k$ denotes 
the realized field at step $k$, $G^k$ denotes its grain graph and $\mathcal S$ denotes the stopping criterion 
for the corresponding analysis.

\begin{algorithm}[htbp]
\caption{VN-GraphGG: the von Neumann-Mullins-based graph grain growth simulation.}
\label{alg:simulation-cycle}
\begin{algorithmic}[1]
\REQUIRE Initial field $L^0$, coefficients $\{M_{\mathrm{eff},g}\}$, time step $\Delta t$, maximum number of steps $K_{\max}$
and stopping criterion $\mathcal S$.
\REQUIRE Fixed graph-to-field reconstruction model $\mathcal D$.
\ENSURE Fields $\{L^k\}$ and grain graphs $\{G^k\}$
\FOR{$k=0,\ldots,K_{\max}-1$}
  \STATE \textbf{\uppercase\expandafter{\romannumeral1}. Graph representation}
  \STATE Construction $G^k$ from $L^k$ and obtain $\{n_g^k, a_g^k, p_g^k\}$.
  \IF{$k>0$ and $\mathcal{S}(L^k,G^k)$ is satisfied}
    \STATE \textbf{Break.}
  \ENDIF

  \STATE \textbf{\uppercase\expandafter{\romannumeral2}. VN update}
  \STATE Compute $\widetilde a_g^{k+1}$ and $\widetilde p_g^{k+1}$ using Eqs.~\ref{eq:discrete-vn} and~\ref{eq:perimeter-target}.
  \STATE Update the corresponding features to form $\widetilde G^{k+1}$.

  \STATE \textbf{\uppercase\expandafter{\romannumeral3}. Reconstruction}
  \STATE Reconstruct a candidate field $L_{\mathrm{cand}}^{k+1}$ using $\mathcal D$ by Eq.~\ref{eq:assignment} from $\widetilde G^{k+1}$.
  \STATE Remove grains smaller than five pixels.
  \STATE Reassign boundary pixels to reduce differences between the measured areas in $L_{\mathrm{cand}}^{k+1}$
  and $\{\widetilde a_g^{k+1}\}$.
  
  \STATE Set the result as $L^{k+1}$.
\ENDFOR
\end{algorithmic}
\end{algorithm}

% =========================================================
\section{Results}

\subsection{Reconstruction fidelity}

We evaluate the trained decoder on a test set with 6,000 graph--field pairs and compare reconstructed 
field with the corresponding ground truth. Fig.~\ref{fig:framework} c shows a representative reconstruction and its
pixelwise classification errors. The mean per-field pixel error is 3.54\%, with a sample standard deviation of 0.22
percentage points, corresponding to a mean pixel accuracy of 96.46\%. Boundary placement and grain adjacency are assessed
using F1 scores, which average 0.916 and 0.973, respectively. The mean absolute area error is 2.04 pixels per grain, 
and the mean missing-grain fraction is 0.020\%. These results support the use of the reconstructed fields to 
supply grain geometry and adjacency for the simulation. Appendix C defines the evaluation metrics, their averaging and 
the associated matching rules.

\subsection{Normal coarsening}

\begin{figure*}[p]
\centering
\includegraphics[width=0.895\textwidth]{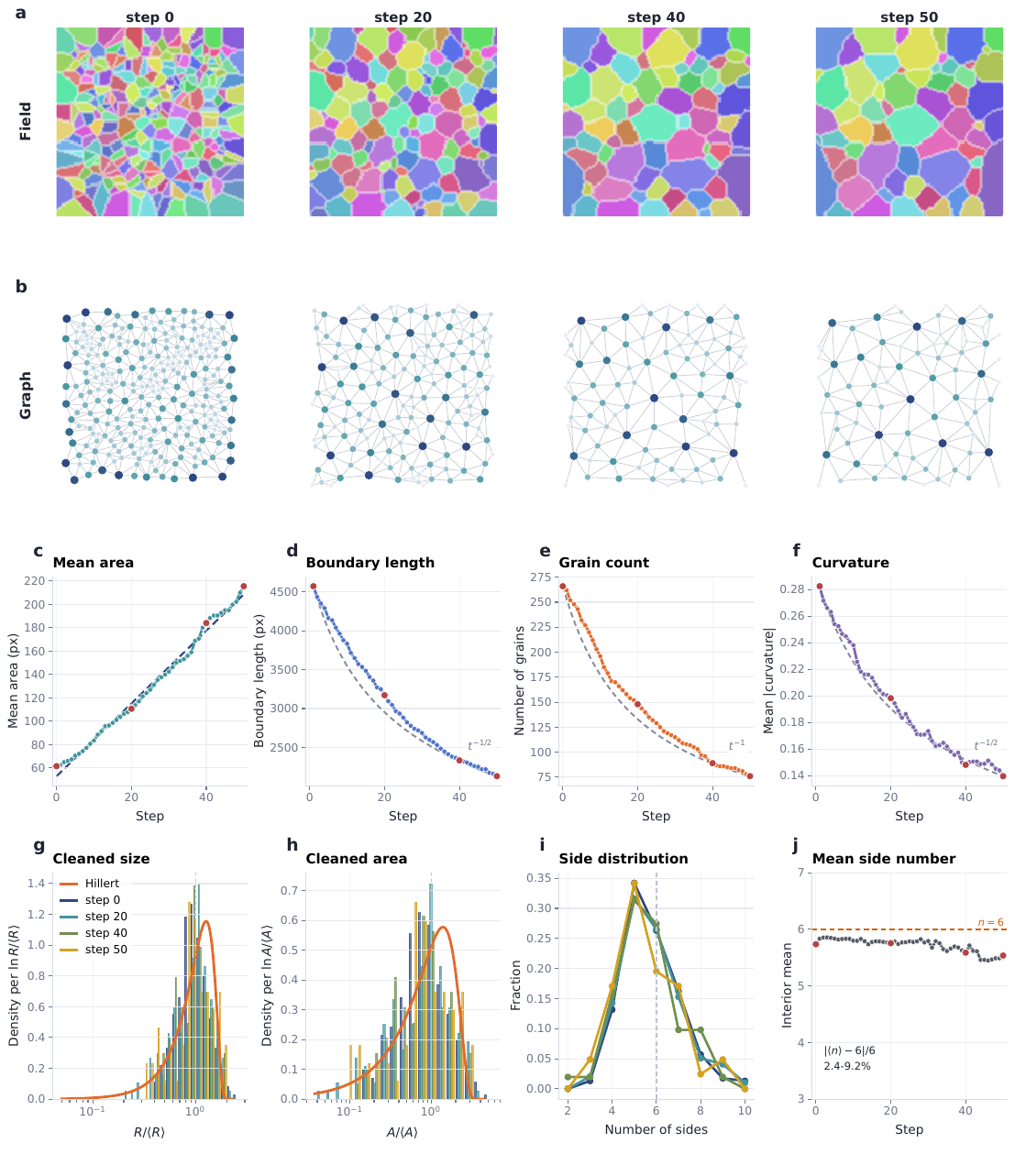}
\caption{\textbf{Normal grain coarsening in a representative simulation.}
(a) Polycrystalline microstructures at steps 0, 20, 40 and 50. (b) Grain graphs without junction nodes are 
presented; node size and color indicate relative grain area. (c--f) Evolution of mean grain area, 
total grain-boundary length, grain count and mean absolute curvature. Red markers identify the displayed steps. 
The dashed line in (c) is a linear fit to the mean-area trajectory; gray dashed curves in (d--f) denote expected 
coarsening trends \cite{Elsey2010}. (g,h) Scaled radius and area distributions, plotted on logarithmic horizontal
axes as densities per unit logarithmic size; orange curves show the Hillert reference distributions \cite{hillert1965}. 
(i) Side-number distributions at the four displayed states. (j) Mean side number of interior grains; 
the dashed line denotes the reference value of six.}
\label{fig:coarsening}
\end{figure*}

We assess whether VN-GraphGG reproduces the statistical behavior of normal grain growth and a representative simulation 
is given in Figure~\ref{fig:coarsening}. We here focus on the population-level coarsening behavior since the method prescribes grain-scale
kinetics without directly resolving grain-boundary motion. 

The microstructure evolves toward fewer, larger grains, as reflected in the corresponding fields and grain graphs (Fig.~\ref{fig:coarsening}a,b). Over 50 steps, the grain count decreases from 266 to 76, while the mean grain area
increases from 61.6 to 2215.6 pixels. The mean grain area increases approximately linearly with simulation step, accompanied by
deceases in total grain boundary length and mean absolute curvature (Fig.~\ref{fig:coarsening}c--f). These trends are
in agreement with the expected signatures of normal coarsening \cite{MULLINS19986219, Elsey2010}. Despite the increase
in mean grain size,the scaled grain area and radius distributions retain broadly similar shapes across the displayed 
states (Fig.~\ref{fig:coarsening}g,h). The side-number distributions also remains similar (Fig.~\ref{fig:coarsening}i). 
The two-sample Kolmogorov--Smirnov tests \cite{Smirnov1939TwoSample} between successive displayed states yields 
$p=0.185-0.798$ for scaled area and $p>0.999$ for side number. The mean side number of interior grains ranges from 
5.45 to 5.85, compared with the ideal value of six (Fig.~\ref{fig:coarsening}j). These observations support the 
reproduction of the statistical trend of normal grain growth. The distributional comparisons are consistent with
self-similarity over the sampled interval.
Appendix C specifies the observable definitions, scaling, exclusions, weighting and statistical comparisons.

\subsection{Prediction of VN kinetics}\label{sec:vn-results}

The simulations with our VN-GraphGG method shows that the grain-area change rate linearly depends on side number, with 
fitted kinetic coefficients close to the prescribed values, confirming that the VN law governs the evolution in the proposed
method. We run 50 simulation for each of two prescribed value, $M_{\mathrm{eff}}=5.0\times10^{-4}$ and $7.5\times10^{-4}$. 
Each simulation contains 50 steps with $\Delta t=0.5$. The normalized area-change rates measured from the resulting fields 
are fitted using ordinary least squares:
\begin{equation}
  \frac{a_g^{k+1}-a_g^k}{\Delta t}=\beta_0+\beta_1(n_g^k-6),
  \label{eq:vn-regression}
\end{equation}
where $n_g^k$ is the side number before the VN update, $\beta_1$ estimates $M_{\mathrm{eff}}$, and $\beta_0$ measures 
offsets at $n_g^k=6$. For each simulation, we fit the observation separately at each step and pool them across all steps. 
Appendix C specifies the observation-selection rules and confidence-interval calculation.

\begin{figure}[p]
\centering
\includegraphics[width=\linewidth]{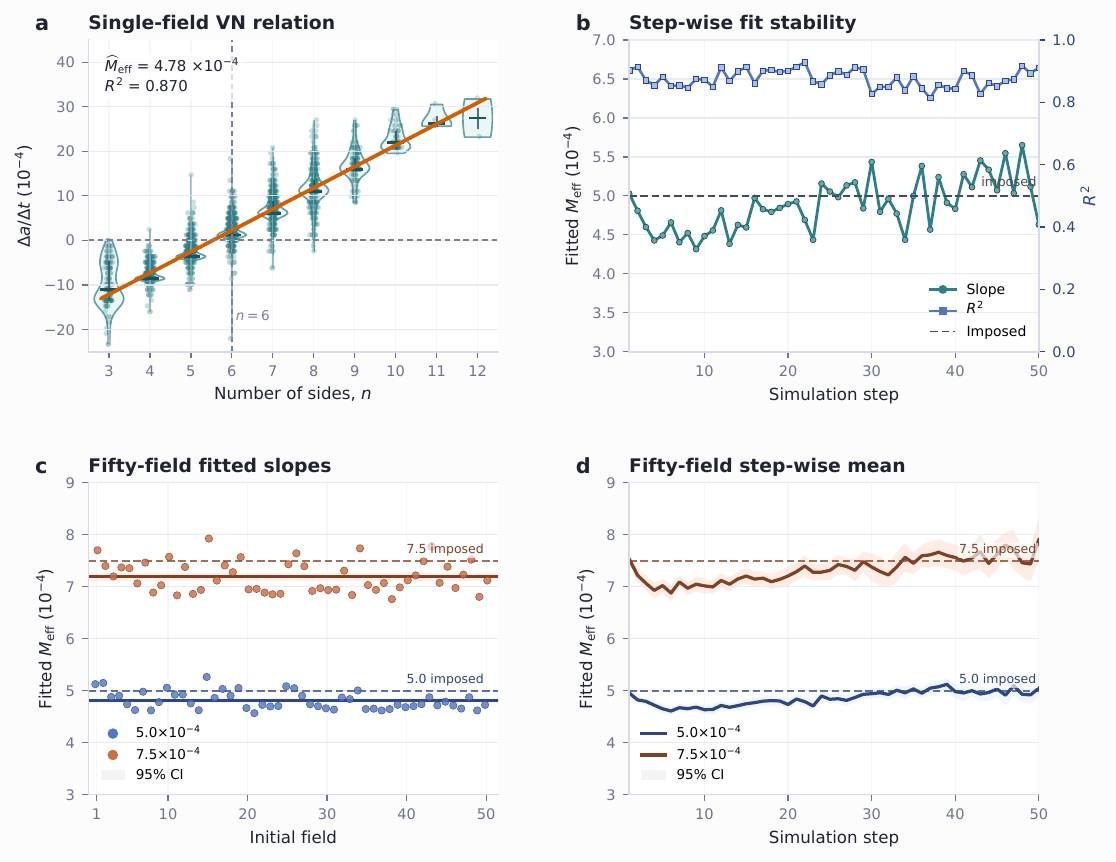}
\caption{\textbf{Prediction of the VN kinetics.} 
(a) Normalized area-change rate $\Delta a/\Delta t$ versus side number in a simulation with 
$M_{\mathrm{eff}}=5.0\times10^{-4}$. Points denote individual observations during the simulation, curved envelopes summarize 
the distribution at each side number and the solid line shows the linear fit. 
(b) Fitted coefficient and coefficient of determination $R^2$ at each step of the same simulation; the dashed line marks
the prescribed coefficient. (c) Coefficients obtained by pooling observations across all steps of each simulation, for
two ensembles of 50 each. Points represent individual simulations, solid lines indicate ensemble average, and shaded bands 
show 95\% confidence intervals of the averages. (d) Fitted coefficients averaged across simulations at each step; 
shaded bands show 95\% confidence interval.}
\label{fig:vn-validation}
\end{figure}

In an illustrative simulation with $M_{\mathrm{eff}}=5.0\times10^{-4}$, the area-change rate remains approximately 
linear in side number (Fig.~\ref{fig:vn-validation}a). We obtain the fitted coefficient of $4.78\times10^{-4}$ with 
$R^2=0.870$ by pooling all observations across steps. The fitted coefficients fitted at each step fluctuate around 
the prescribed value (Fig.~\ref{fig:vn-validation}b), showing that the area-rate dependence on side number is maintained
during the simulation.

For simulations with two prescribed coefficient values of 
$M_{\mathrm{eff}}=5.0\times10^{-4}$ and $M_{\mathrm{eff}}=7.5\times10^{-4}$, the fitted coefficients
form two distinct groups around their respective prescribed values (Fig.~\ref{fig:vn-validation}c). The ensemble-averaged 
coefficients are $4.808\times10^{-4}$ and $7.190\times10^{-4}$, respectively, which are 3.8\% and 4.1\% below the prescribed 
coefficient. The fitted coefficients at each step across simulations with the same setting track its prescribed value 
across the trajectory (Fig.~\ref{fig:vn-validation}d). All these results show that the VN law controls the grain evolution in
the proposed method, and the prescribed coefficient produces the corresponding change in the dependence between grain
area change rates and side number.

\subsection{Phase-field comparison and computational efficiency}

VN-GraphGG results shows bulk coarsening trends similar to those simulations of a phase-field method (PFM) implemented in the MOOSE framework \cite{permann2020}. The proposed method also gains a significantly reduction of computational time in a five-case benchmark over PFMs. 

\begin{figure}[p]
\centering
\includegraphics[width=0.90\textwidth]{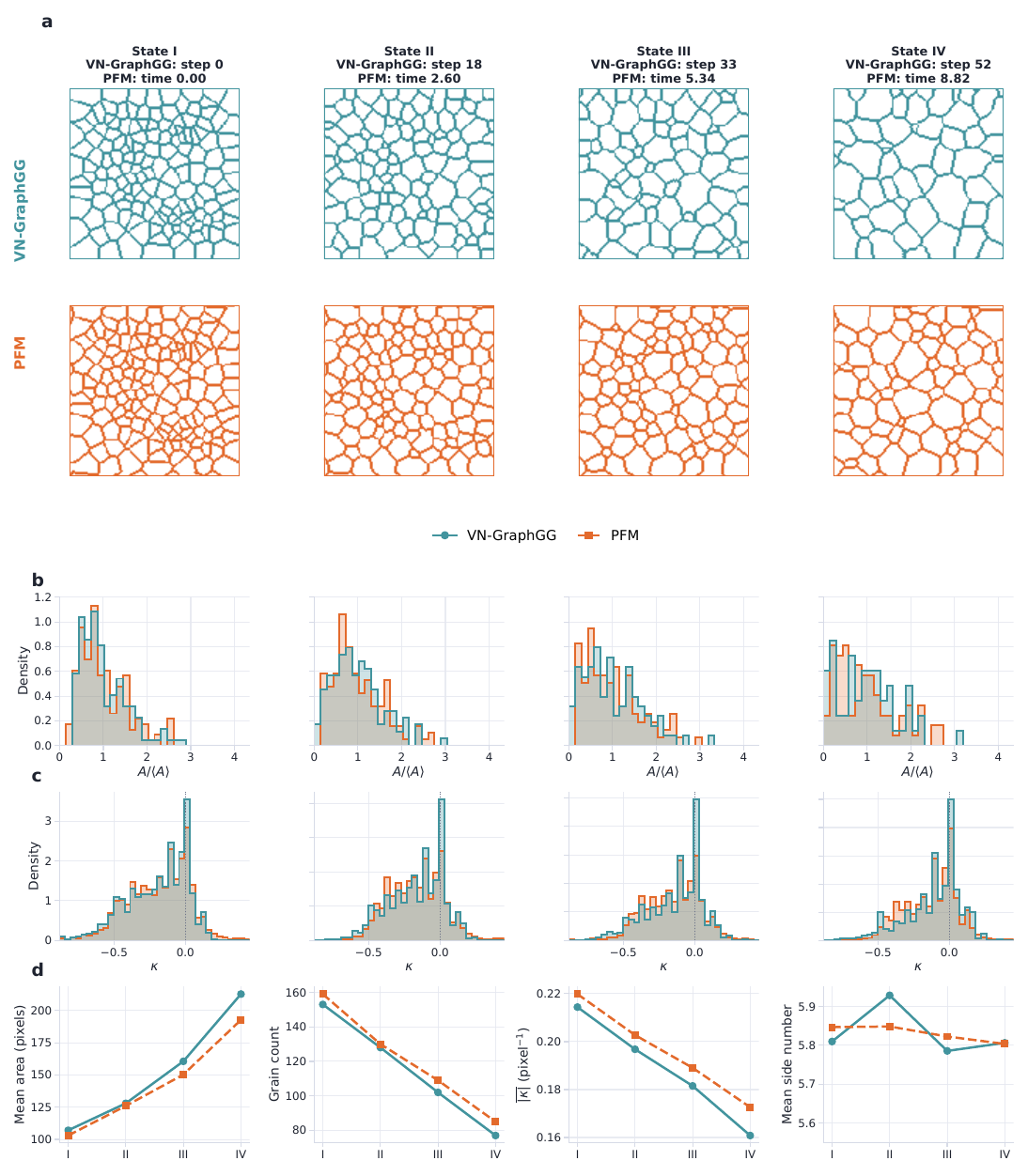}
\caption{\textbf{Comparison between VN-GraphGG and phase-field coarsening.} (a) Microstructures at four states 
with comparable grain counts. VN-GraphGG step number and PFM simulation time are shown separately. (b) Scaled
grain-area distributions at these states. (c) Curvature distributions. (d) Mean grain area, grain count, 
boundary-pixel-weighted mean absolute curvature $\overline{|\kappa|}$ and mean side number of interior grains at
the selected states.}
\label{fig:phasefield-comparison}
\end{figure}

In an example 
shown in Fig.~\ref{fig:phasefield-comparison}, states are paired at comparable grain counts to represent similar
stages of coarsening. Both methods show increasing mean grain area, decreasing mean absolute curvature, and a mean side 
number of interior grains near six (Fig.~\ref{fig:phasefield-comparison} d). The scaled area distributions and 
curvature distributions also show broadly similar shapes (Fig.~\ref{fig:phasefield-comparison} b and c). These
observations support agreement in bulk coarsening behavior at the selected states.
Appendix D specifies the initialization, boundary conditions,
observables, and solver settings.

We compare computational cost across five cases at the grain-count threshold $N_{\mathrm{tar}}=N^0/2$, where $N^0$ is the 
initial grain count. This threshold provides a common measure of coarsening progress for methods with different time setting. 
The calculation of wall time is detailed in Appendix E. 
Both methods are implemented on a AMD Ryzen Threadripper PRO 3955WX CPU with one CPU thread. Across the five cases, the 
mean wall time to the threshold is 38.43~s for VN-GraphGG method and 5,247.01~s, or about 1.5~h, for the phase-field method (Table~\ref{tab:benchmark}). The arithmetic mean of the case-wise 
PFM-to-VN-GraphGG wall-time ratios is 136.9$\times$. The wall-time of VN-GraphGG method is averaged over three repeated runs for
each case, whereas each PFM case is run once.

\begin{table*}[!t]
\centering
\caption{\textbf{Wall time to the half-grain-count threshold for the VN-GraphGG and PFM.}
VN-GraphGG wall times are reported as mean value and the standard deviation over three repeated runs; 
each PFM case is performed once. Threshold wall times are linearly interpolated from the recorded 
grain-count series. The Mean column reports arithmetic means across the five cases.}
\label{tab:benchmark}
\begingroup
\small
\setlength{\tabcolsep}{4.4pt}
\renewcommand{\arraystretch}{1.08}
\resizebox{\textwidth}{!}{
\begin{tabular}{@{}lrrrrrr@{}}
\toprule
& \multicolumn{5}{c}{Cases} & \\
\cmidrule(lr){2-6}
Quantity & 11967 & 13406 & 18078 & 28383 & 32613 & Mean \\
\midrule
\multicolumn{7}{@{}l}{\textit{Grain counts}} \\
\quad Initial grains & 153 & 93 & 145 & 254 & 190 & -- \\
\quad Target grains & 76.5 & 46.5 & 72.5 & 127.0 & 95.0 & -- \\
\quad Final grains, Graph & 75.0 & 46.0 & 71.0 & 124.0 & 95.0 & -- \\
\quad Final grains, PFM & 76.0 & 46.0 & 72.0 & 127.0 & 95.0 & -- \\
\addlinespace[0.4em]
\multicolumn{7}{@{}l}{\textit{Numerical steps}} \\
\quad VN-GraphGG & 48.0 & 73.0 & 46.0 & 23.0 & 31.0 & -- \\
\quad PFM & 92 & 126 & 105 & 44 & 66 & -- \\
\addlinespace[0.4em]
\multicolumn{7}{@{}l}{\textit{Wall time}} \\
\quad VN-GraphGG (s) & 41.74$\pm$0.07 & 51.23$\pm$1.01 & 38.40$\pm$0.67 & 28.09$\pm$0.06 & 32.69$\pm$0.60 & 38.43 \\
\quad PFM (s) & 5,525.50 & 6,504.00 & 6,065.58 & 3,624.84 & 4,515.14 & 5,247.01 \\
\quad Speedup & 132.4$\times$ & 126.9$\times$ & 158.0$\times$ & 129.1$\times$ & 138.1$\times$ & \textbf{136.9$\times$} \\
\addlinespace[0.4em]
\multicolumn{7}{@{}l}{\textit{Mean wall time per step}} \\
\quad VN-GraphGG (s) & 0.870 & 0.702 & 0.835 & 1.221 & 1.055 & 0.936 \\
\quad PFM (s) & 60.060 & 51.619 & 57.767 & 82.383 & 68.411 & 64.048 \\
\quad Speedup & 69.1$\times$ & 73.5$\times$ & 69.2$\times$ & 67.5$\times$ & 64.9$\times$ & \textbf{68.8$\times$} \\
\bottomrule
\end{tabular}
}
\endgroup
\end{table*}

\subsection{Abnormal grain growth}\label{sec:agg-results}

We apply VN-GraphGG to study abnormal grain growth and investigate the effects of kinetic heterogeneity on AGG \cite{decost2017,liu2019,frazier2015,marvel2022}. The trained reconstruction model is fixed across all conditions.
Each grain initially receives a baseline effective kinetic coefficient is $M_{\mathrm{eff}}=5.0\times10^{-4}$. We then 
randomly sample $N_{\mathrm{selected}}$ grains from the $N_{\mathrm{eligible}}$ initial grains that do not touch the 
domain boundary and assign them with a higher coefficient $M_{\mathrm{high}}$. 

The selected fraction is $f_{\mathrm{high}}=N_{\mathrm{selected}}/N_{\mathrm{eligible}}$. The kinetic-coefficient 
contrast is $\chi=M_{\mathrm{high}}/M_{\mathrm{eff}}$. The fraction sweep varies $f_{\mathrm{high}}$ from 1\% to 100\% at 
fixed contrast $\chi=2$. The contrast sweep varies $\chi$ from 1 to 3 at $f_{\mathrm{high}}=10\%$. Each condition comprises 
30 simulations. In all simulations, $\Delta t=0.5$ and $K_{\max}= 60$.

AGG outcomes are evaluated before that the grain count reaches or falls below half its initial value, i.e.,
steps $k\leq k_{50}=\min\{k:N^k\leq \frac{N^0}{2}\}$. 
For a selected grain $i$ with neighborhood $\mathcal N_i^k$ at $k$ step, the local area ratio is defined as
\begin{equation}
  R_i^k=\frac{a_i^k}{\operatorname{median}\{a_j^k:j\in\mathcal N_i^k\}}.
  \label{eq:agg-ratio}
\end{equation}
A grain is classified as exhibiting sustained AGG if $R_i^k\geq 5$ at three consecutive steps.

\begin{figure}[!ht]
\centering
\includegraphics[width=0.95\textwidth]{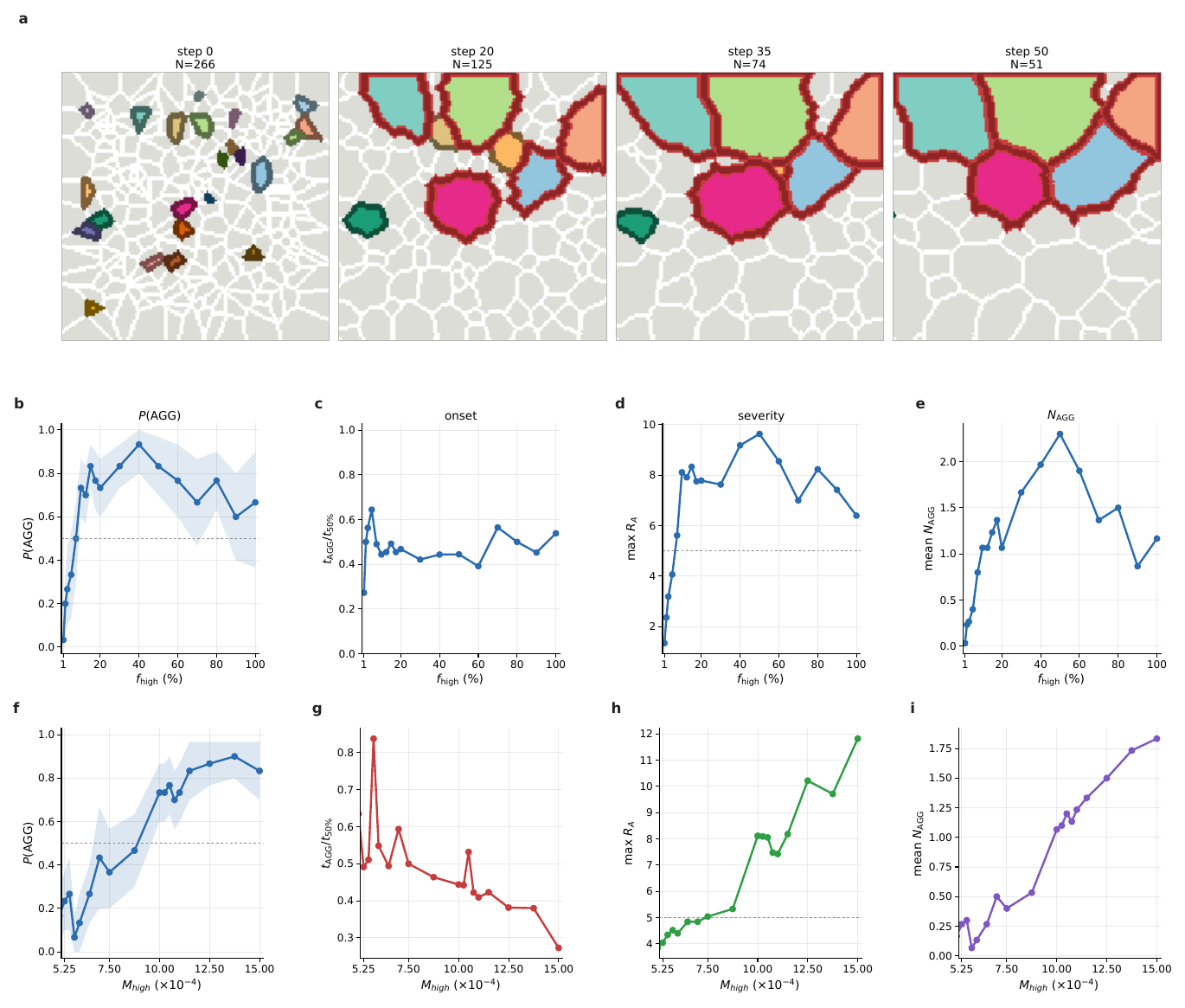}
\caption{\textbf{Effects of kinetic heterogeneity on abnormal grain growth.}
(a) A representative simulation with $f_{\mathrm{high}}=10\%$ and $\chi=2$. 
Unselected grains are shown in gray, selected high-coefficient grains are colored, and red boundary outlines mark selected
grains that satisfy the criterion for sustained AGG during the simulation. (b--e) High-coefficient-fraction sweep with
$1\%\leq f_{\mathrm{high}} \leq100\%$ and $\chi=2$. The panels show (b) AGG occurrence fraction,
(c) mean normalized onset time, (d) mean severity and (d)mean AGG grain count. (f--i) Coefficient-contrast sweep with 
$f_{\mathrm{high}}=10\%$ and $1\leq \chi\leq 3$. Shaded bands in (b) and (f) show pointwise 95\% percentile intervals 
from resampling.}
\label{fig:agg-application}
\end{figure}

For each condition, we report the AGG occurrence fraction $P(AGG)$, normalized onset time $\frac{t_{AGG}}{t_{50}}$, 
severity $\textit{max}\, R_A$ and AGG grain count $N_{AGG}$ here. The occurrence $P(AGG)$ is the proportion of simulations 
containing at least one AGG grain. The onset time is the earliest step qualifying the criterion, denoted by $t_{AGG}$, and 
is normalized by $t_{50}$. The normalized onset time is summarized among simulations in which AGG happens. Severity
$\textit{max}\, R_A$ is the maximum local area ratio $R_i^k$ over all grains $i$ and steps $k$. The AGG count $N_{AGG}$ is
the number of grains satisfying the AGG criterion. Figure~\ref{fig:agg-application} b-i show the occurrence, the median
normalized onset time, the median severity, and the mean AGG grain count for both fraction sweep and contrast sweep.
Appendix E provides the complete definitions and resampling procedure for analysis.

From Fig.~\ref{fig:agg-application}a, we can find that assigning a higher coefficient alone does not guarantee abnormal growth. Under the VN law, the coefficient 
multiplies $(n_g-6)$, accelerating growth for $n_g>6$ and shrinkage for $n_g<6$. In this representative simulation, 
many selected grains disappear or remain comparable in size to their neighbors. Only a subset survives and meets 
the sustained-AGG criterion (Fig.~\ref{fig:agg-application}a). The outcome therefore depends on both the kinetic 
coefficient and the evolving neighborhood, which is qualitatively consistent with experimental observations \cite{conditionsAbnormalGrainGrowth2012,Dressler1996, decost2017,liu2019}.

At a fixed $\chi=2$, the highest observed AGG occurrence fraction is 0.93 around $f_{\mathrm{high}}=40\%$, compared with 
0.03 at $f_{\mathrm{high}}=1\%$. With increasing fraction, the occurrence fraction subsequently declines to 0.67 at $f_{\mathrm{high}}=100\%$ 
(Fig.~\ref{fig:agg-application}b). Mean severity and the mean AGG grain count also reach their highest observed values
around an intermediate fraction, whereas mean normalized onset has no clear monotonic trend (Fig.~\ref{fig:agg-application}c--e). 

With a fixed $f_{\mathrm{high}}=10\%$, increasing $\chi$ produces a different response. The occurrence nearly increases 
from 0.17 $\chi=1.05$ ($M_{\mathrm{high}}=5.25\times 10^{-4}$) to 0.83 at $\chi=3$ ($M_{\mathrm{high}}=15\times 10^{-4}$) 
(Fig.~\ref{fig:agg-application}f). The mean normalized onset time decreases from 0.63 to 0.27 (Fig.~\ref{fig:agg-application}g).
The mean severity and mean AGG grain count also generally increase from 3.84 to 11.82, and 0.17 to 1.83, respectively 
(Fig.~\ref{fig:agg-application}h and i).

These controlled sweeps show that the distribution and magnitude of the kinetic coefficients influence abnormal grain
growth in different ways. Varying the selection fraction $f_{\mathrm{high}}$ produces a non-monotonic population-level
response, whereas increasing the contrast $\chi$ generally increases AGG occurrence and severity.

% =========================================================
\section{Conclusions and Discussion}

We develop VN-GraphGG for two-dimensional grain-growth simulation by combining the von Neumann--Mullins law 
with neural graph-to-field reconstruction. In this method, we assign distinct roles to the physical law and 
machine learning. The VN law is applied to control grain-area increments on a grain graph, while the neural 
model is trained to reconstruct the polycrystalline microstructure from the graph. We then apply postprocessing 
to evolve the topology of grain boundary network. This separation of grain-area evolution and geometric reconstruction 
allows to adjust the kinetic coefficients for evolution independently and no retraining is needed for the neural 
reconstruction model. Our method leverages grain-scale kinetics and graph neural networks to achieve efficient 
simulation and enables ensemble simulations with a range of kinetic conditions.

We first validate VN-GraphGG through reconstruction fidelity, normal coarsening behavior and prediction of prescribed 
kinetics. The reconstruction tests support the use of the reconstructed fields to supply the geometry and adjacency 
needed for subsequent steps. The simulated coarsening statistics retain the principal population-level trends 
expected for normal grain growth. The area-change rates in simulations remain approximately 
linear in side number and distinguish the two prescribed coefficients. These results show that 
VN-GraphGG can provide spatial evolution of microstructure governed by the VN law.

The proposed method complements both learned evolution surrogates and statistical VN-based models. 
The prescribed grain-scale kinetics remain separate from the neural reconstruction model, eliminating the need for 
classical simulations to generate training datasets, as required by evolution surrogates. The evolution also 
can be varied by the kinetic coefficients without retraining the graph decoder. Meanwhile, the method
takes advantage of the graph representation of grains to reconstruct the polycrystalline microstructure.
Topological changes therefore are incorporated through the reconstruction and postprocessing, avoiding a 
separate topological-event solver. This separation enables pathways for simulating grain growth with grain-scale kinetics.

VN-GraphGG reaches the half-grain-count threshold with a significant reduction of computational cost compared with
phase field methods, which is mainly due to the efficient grain-scale kinetics and graph neural networks. The VN law 
is used to control grain area evolution without the introduction of phase field equations. The graph convolution 
decoder is applied to reconstruct the microstructure that the area evolution alone does not determine. This combination 
makes ensemble simulation over a range of kinetic conditions computationally practical.

We apply VN-GraphGG to investigate the effects of kinetic heterogeneity on grain growth. In the fraction sweep, we find 
that the occurrence fraction peaks at an intermediate high-coefficient grain fraction and then decreases as the selected 
fraction increases. The subsequent decrease may reflect stronger competition among high-coefficient grains as their 
fraction increases, limiting the development of size differences relative to their neighbors. At the fixed fraction, the 
occurrence of AGG generally increase as the coefficient contrast increases. These findings distinguish effects of the 
magnitude and the distribution of kinetic heterogeneity on grain growth, and provide a basis for investigating how their 
influence could be used to control abnormal grain growth.

VN-GraphGG provides an efficient two-dimensional tool for ensemble microstructure evolution with explicit
control of grain-scale kinetics. Its separation of the prescribed kinetics from spatial reconstruction offers a natural basis
for future extensions that incorporate larger domains, additional microstructure classes
and experimental fields. Anisotropic boundary properties, solute drag, second-phase pinning and internal stress
could be introduced through corresponding kinetic formulations and input features
\cite{Rohrer2023,qiuWhyGrainGrowth2025}. In three dimensions, the generalized MacPherson--Srolovitz relation provides
a possible kinetic basis for extending the same graph-to-field architecture \cite{macphersonNeumannRelationGeneralized2007}. 
These directions offer a path toward efficient grain-growth simulations
that incorporate a broader range of physical mechanisms through grain-scale kinetics.

% =========================================================
%\subsection*{Data availability}

%The generated field collections, benchmark records and processed measurements will be deposited in a versioned public
%repository with a persistent identifier upon acceptance and before publication.
%% Before final acceptance, replace this sentence with the repository name, DOI %and license.

%\subsection*{Code availability}

%The custom code will be released in a versioned public repository upon acceptance and archived with a persistent
%identifier before publication. Until then, it will be available confidentially to editors and reviewers upon request.
%% Before final acceptance, add the repository URL, archive DOI, license and environment specification.

\section*{Acknowledgements}

The work of Y.X. was supported by the National Key Research and Development Program of China No.~2025YFA1016800 and the
Hong Kong Research Grants Council Collaborative Research Fund C7140-25G. The work of S.D. was support  by the
National Natural Science Foundation of China No.~12471409.

%\section*{Author contributions}

%Y.H.: Conceptualization, Methodology, Investigation, Writing - original draft.
%Z.L.: Conceptualization, Methodology, Investigation, Writing - review and editing.
%S.D.: Conceptualization, Methodology, Investigation, Writing - review.
%C.M.: Conceptualization, Methodology, Writing - review.
%Y.X.: Conceptualization, Methodology, Investigation, Supervision, Writing - review and editing.

\section*{Competing interests}

The authors declare no competing interests.

% =========================================================

\bibliography{sample_paper}

%\clearpage
\appendix
\makeatletter
\@addtoreset{figure}{section}
\@addtoreset{table}{section}
\makeatother
\section{Block \uppercase\expandafter{\romannumeral1}: Heterogeneous grain graph construction}\label{app:graph}

The modular takes a $128\times 128$ microstructure field with $N^0$ grains as input. An integer number $I$
between 1 and $N^0$ is assigned to each grain and is called the grain index. $I(x)$ denote the grain index at 
pixel $x$. The field $L$ then is a $128\times 128$ matrix consisting of $I(x)$. It is vectorized and converted 
into a heterogeneous graph with grain nodes and junction nodes as follows:
\begin{enumerate}
\item \textbf{Create grain nodes.} Create one node for each distinct grain index and position it at the centroid
of all pixels with that index.
\item \textbf{Determine grain--grain edges.} Scan horizontally and vertically adjacent pixel pairs. If their grain
indices are distinct, connect the corresponding grain nodes, retaining one undirected edge per
unordered grain pair. Diagonal contacts and contacts across opposite domain boundaries do not create edges.
\item \textbf{Create junction nodes.} Examine interior pixels whose index differs from at least one of their four
horizontal or vertical neighbors. Create a junction node at the pixel coordinates if its $3\times3$
neighborhood contains at least three distinct grain indices \cite{qin2023grainnn, Yang2024TripleJunction}.
\item \textbf{Determine grain--junction edges.} Count the pixels belonging to each grain in the $3\times3$
neighborhood of junctions. Connect the junction to the grain nodes corresponding to the three most frequent indices. 
\item \textbf{Prepare for input} Assemble the node features listed in Table~\ref{tab:features} and represent each
undirected edge in both directions for message passing. This gives grain-to-grain and grain-to-junction relations.
\end{enumerate}

The heterogeneous grain graph comprises 13 features for grain nodes and nine features for junction nodes 
(Table~\ref{tab:features}). All areas, perimeters and other coordinate features are normalized 
by $128^2$, $4\times128$, and 128, respectively.
\begingroup
\small
\setlength{\tabcolsep}{4pt}
\begin{longtable}{@{}p{0.19\textwidth}p{0.15\textwidth}p{0.59\textwidth}@{}}
\caption{Input features of the grain graph.}\label{tab:features}\\
\toprule
Feature group & Dimension & Feature descriptions \\
\midrule
\endfirsthead
\toprule
Feature group & Dimension & Feature descriptions \\
\midrule
\endhead
Grain & 13 & Normalized area and perimeter; normalized centroid; normalized adjacency degree and 
an indicator of whether the degree is at least three; normalized junction count and mean junction count; 
mean and standard deviation of distances from the grain centroid to junction; log-transformed 
junction count; sine and cosine of orientation. \\
Junction candidate & 9 & Normalized position; connected-grain count; means and standard deviations of the areas and
perimeters of connected grains; mean and standard deviation of distances to connected-grain centroids. \\
\bottomrule
\end{longtable}
\endgroup

\section{\uppercase\expandafter{\romannumeral3}: Graph-to-field reconstruction}
\label{app:architecture}

\subsection{Architecture of the graph decoder}
%Structural and connected-component grain features are encoded separately by two-layer multilayer perceptrons (MLPs)
%with rectified linear unit (ReLU) activations and a hidden width of 512. Let $\mathbf h_g^{\mathrm{s}}$ and
%$\mathbf h_g^{\mathrm{c}}$ denote the structural and connected-component representations, respectively. The
%connected-component contribution is gated before being added to the structural representation:
%\begin{equation}
%\mathbf h_g^{0}=\mathbf h_g^{\mathrm{s}}+
%\sigma(\mathbf W_{\mathrm{gate}}\mathbf h_g^{\mathrm{c}})
%\odot\tanh(\mathbf W_{\mathrm{comp}}\mathbf h_g^{\mathrm{c}}),
%\end{equation}
%where $\sigma$ is the logistic function, $\odot$ denotes element-wise multiplication, and
%$\mathbf W_{\mathrm{gate}}$ and $\mathbf W_{\mathrm{comp}}$ are learned projection matrices.

In the proposed graph convolutional decoder, the heterogeneous grain graph $G$ are firstly 
passed into two-layer multilayer perceptrons (MLPs), which lift the features on grain and junctions 
into two 512-dimensional latent vectors separately, with a rectified linear unite (ReLu) activation 
after each fully-connected layer.

The latent vectors then are input into three heterogeneous graph convolution layers, which are connected 
by residual. Within each layer, a separate graph convolution is applied to each edge type and messages from
neighbors are summed at each node. A ReLU activation is applied to the sum, and the result is added to the
vectors from the preceding layer.

Two-layer post-message-passing MLPs are then applied separately to process the latent vectors. 
For each grain, the latent vectors of the connected junction nodes are averaged to form a junction-informed vector, 
ensuring that each grain has a vector with the features on grain nodes and another vector with the features 
on connected junction nodes. These two latent vectors for each grain then are concatenated and mapped to 
a 192-dimensional grain query vector $\mathbf q_g$ and a scalar bias $b_g$ by a fusion layer.

At the same time, a $128\times 128$ field template gives normalized coordinates $\mathbf x=(x_1,x_2)\in[0,1]^2$.
The coordinates are augmented to position vectors with sine and cosine features at eight fixed Fourier 
frequency $2^k$, $k=0,\ldots,7$ for each pixel $\mathbf{x}$:
\begin{equation}
\boldsymbol\varphi(\mathbf x)=
\left[\mathbf x,
\left\{\sin(2\pi2^k x_d),\cos(2\pi2^k x_d)\right\}_{d\in\{1,2\},\,k=0}^{7}
\right].
\end{equation}
This Fourier positional encoding provides slowly and rapidly varying spatial features to help the coordinate 
MLP represent grain shapes and boundary details. The resulting 34-dimensional vector for each pixel $\mathbf{x}$ 
is mapped by a two-layer MLP to a 192-dimensional position vector $\mathbf p_x$. 

For each grain $g$, the decoder defines the score at each pixel $\mathbf{x}$ as,
$$s_{xg}=p_x^{\mathsf T}\mathbf q_g+b_g,$$
and the grain with the largest score determines the classification of the pixel (Eq~\ref{eq:assignment}).

%\subsection{Training and model selection}
%\label{app:training}

\subsection{Datasets}

The dataset contains 40,000 synthetic Voronoi fields represented on $128\times 128$ grids. Each field is a Voronoi tessellation,
with one integer index assigned to every grain. The heterogeneous grain graphs are extracted directly from those fields using
the procedure in Section~\ref{app:graph}. The graphs serve as the input to the decoder, and their corresponding fields are ground truths
for reconstruction. In constructing the dataset, a field is accepted only if it contains at least eight grains and at least one junction.

Every field-graph pair is assigned to an identifier and we shuffle those identifiers using a NumPy random-number generator 
with seed 42, to partition the dataset into training, validation and test subsets. Those field-graph pairs are divided in a 70:15:15 ration, giving 28,000 training data, 6,000 validation data and 6,000 test data.

\subsection{Training history}

\begin{figure}[htbp]
\centering
\includegraphics[width=\textwidth]{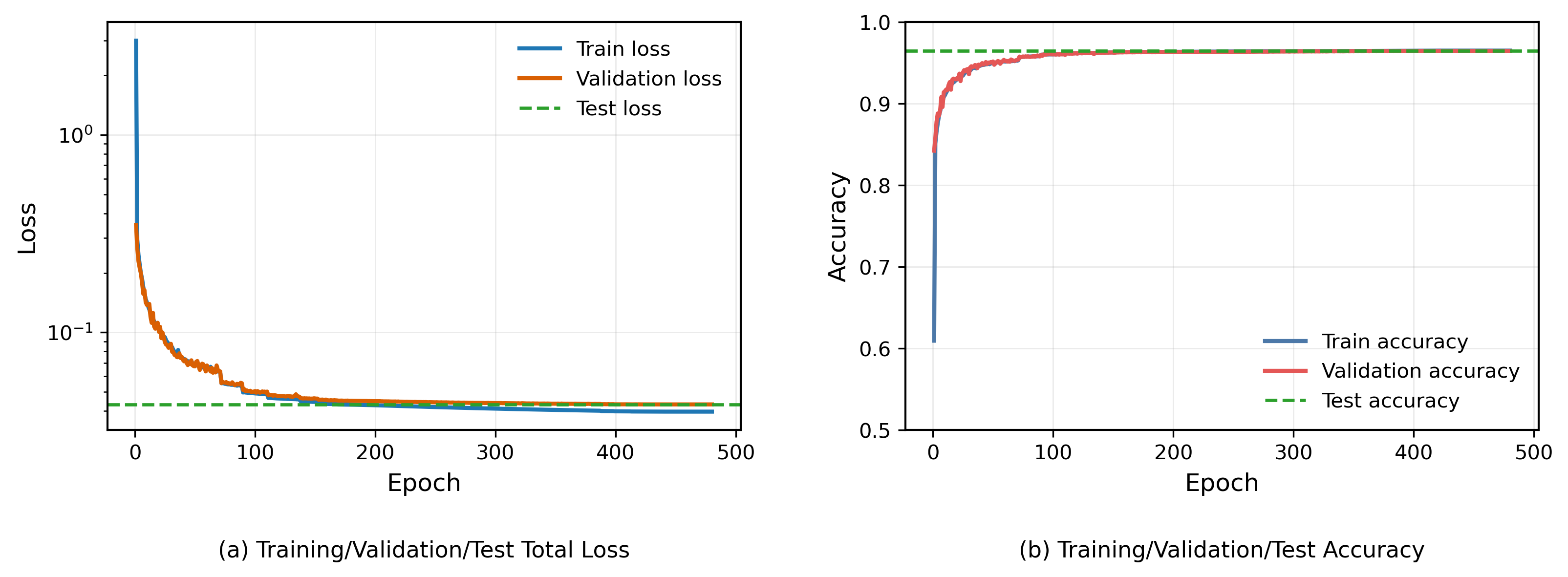}
\caption{Training history of the decoder. (a) Total loss and (b) pixel-classification accuracy over 480 epochs. Blue and
orange curves show training and validation results, respectively. Green dashed lines indicate the test values.}
\label{fig:training-curves}
\end{figure}

The training history of the decoder is given in \ref{fig:training-curves}.

\subsection{Optimization and checkpoint selection}

The decoder is trained by minimizing the total loss function using Adam. We use mini-batches of 16 field--graph pairs, an 
initial learning rate of $5\times 10^{-4}$ and a weight-decay coefficient of $1.0\times 10^{-5}$. The global gradient norm 
is clipped at 5.0.

The total loss function is evaluated on the validation set after every epoch. A reduce-on-plateau schedule monitors 
the validation loss function and multiplies the learning rate by 0.5 when no improvement is observed for five consecutive 
epochs. Training is allowed to proceed for at most 500 epochs and is terminated after 20 consecutive epochs without improvements. 
The checkpoint with the lowest validation objective is used for testing and all simulations reported in the 
main manuscript. All architecture and training settings are given in Table~\ref{tab:hyperparameters}.

\begin{table}[htbp]
\centering
\caption{Settings of the decoder.}
\label{tab:hyperparameters}
\small
\begin{tabular}{@{}p{0.64\textwidth}p{0.31\textwidth}@{}}
\toprule
Setting & Value \\
\midrule
Input dimensions (structural / junction) & 13 / 9 \\
Hidden width & 512 \\
Heterogeneous graph-convolution stages & 3 \\
Pre- and post-message-passing MLP depths & 2 / 2 \\
Pixel-vector and grain-query dimensions & 192 \\
Fourier frequency bands & 8 \\
Dropout & 0 \\
Trainable parameters & 8,304,577 \\
Training / validation / independent test sets & 28,000 / 6,000 / 6,000 \\
Mini-batch size & 16 \\
Initial learning rate & $5.0\times10^{-4}$ \\
Weight decay & $1.0\times10^{-5}$ \\
Learning-rate scheduler patience / reduction factor & 5 / 0.5 \\
Gradient-norm limit & 5.0 \\
Focal-loss exponent $\gamma$ & 2.0 \\
Relative soft grain-area loss weight & 10.0 \\
\bottomrule
\end{tabular}
\end{table}

After training and checkpoint selection, the obtained decoder is frozen. The test subsets is used only 
for evaluation and no parameters will be updated at this stage. The decoder after evaluation on reconstruction
fidelity is applied to simulating grain growth.

\section{Evaluation metrics and statistical analysis}
\label{app:evaluation}

\subsection{Reconstruction fidelity}

For each field in the test set, let $\Omega$ denote the
$128\times128$ pixel domain, $I(x)$ the reference grain index at pixel $x$,
and $\widehat{I}(x)$ the grain index assigned by the reconstruction decoder. Define
the reference and reconstructed grain sets as
\begin{equation}
\mathcal G=\{I(x):x\in\Omega\},
\qquad
\widehat{\mathcal G}=\{\widehat{I}(x):x\in\Omega\},
\end{equation}
with grain counts $N=|\mathcal G|$ and
$\widehat N=|\widehat{\mathcal G}|$, respectively.

Pixel accuracy measures pointwise accuracy:
\begin{equation}
\operatorname{Acc}_{\mathrm{pix}}
=\frac{1}{|\Omega|}\sum_{x\in\Omega}\mathbf{1}\!\left[\widehat{I}(x)=I(x)\right],
\end{equation}
where $\mathbf{1}[\cdot]$ is the indicator function. For each reference grain
$g\in\mathcal G$, its reference and reconstructed areas, measured in pixels,
are
\begin{equation}
A_g=\sum_{x\in\Omega}\mathbf{1}[I(x)=g],
\qquad
\widehat A_g=\sum_{x\in\Omega}\mathbf{1}[\widehat{I}(x)=g].
\end{equation}
The mean absolute grain-area error for one field is
\begin{equation}
\varepsilon_A
=\frac{1}{N}
\sum_{g\in\mathcal G}\left|\widehat A_g-A_g\right|.
\end{equation}
The average is taken over the reference grains. If a reference grain is absent from the reconstructed field,
then $\widehat A_g=0$ and its full reference area contributes to $\varepsilon_A$.

Grain retention is further quantified by the absolute grain-count error and
the missing-grain fraction:
\begin{equation}
\varepsilon_N=\left|\widehat N-N\right|,
\qquad
f_{\mathrm{miss}}
=
\frac{\varepsilon_N}{N}.
\end{equation}

To evaluate interfaces and grain contacts, let $\mathcal P_4$ be the set of
unordered pairs of adjacent pixels. The boundary pixels in the field are defined as
\begin{equation}
\mathcal B(I)
=\left\{x\in\Omega:\exists\,\{x,y\}\in\mathcal P_4
\ \text{such that}\ I(x)\neq I(y) \right\},
\end{equation}
and its grain-adjacency set is
\begin{equation}
\mathcal E(I)
=\left\{\{I(x),I(y)\}:\{x,y\}\in\mathcal P_4,\ I(x)\neq I(y)
\right\}.
\end{equation}
Thus, $\mathcal E(I)$ contains one unordered pair for every pair of distinct
grain indices sharing a horizontal or vertical boundary.

For reconstructed and reference sets $S_{\mathrm{rec}}$ and
$S_{\mathrm{ref}}$, precision, recall and their harmonic mean are
\begin{equation}
\operatorname{Prec}
=\frac{|S_{\mathrm{rec}}\cap S_{\mathrm{ref}}|}
{|S_{\mathrm{rec}}|},
\qquad
\operatorname{Rec}
=\frac{|S_{\mathrm{rec}}\cap S_{\mathrm{ref}}|}
{|S_{\mathrm{ref}}|},
\qquad
F_1=\frac{2\,\operatorname{Prec}\operatorname{Rec}}
{\operatorname{Prec}+\operatorname{Rec}}.
\end{equation}
The reported boundary and adjacency scores are therefore
\begin{equation}
F_1^{\mathrm{bnd}}
=
F_1\!\left(\mathcal B(\widehat{I}),\mathcal B(I)\right),
\qquad
F_1^{\mathrm{adj}}
=
F_1\!\left(\mathcal E(\widehat{I}),\mathcal E(I)\right).
\end{equation}
Boundary matching is exact: a reconstructed boundary pixel is counted as
correct only when the reference field contains a boundary pixel at the same
location. No spatial dilation or distance tolerance is used for the reported
boundary score.

Every metric is first calculated separately for each field. If $m^{(L)}$
denotes a field-level metric for the field $L$, the reported dataset mean is
\begin{equation}
\overline m
=
\frac{1}{N_{\mathrm{eval}}}
\sum_{L}^{N_{\mathrm{eval}}}m^{(L)},
\qquad
N_{\mathrm{eval}}=6000.
\end{equation}
Accordingly, all fields receive equal weight. In particular,
$\varepsilon_A$ is first averaged over reference grains within a field and
then averaged over fields.

\subsection{Prediction of VN kinetics}

The prescribed area $\widetilde a_g^{k+1}$ satisfies the discrete VN update
by construction. The VN kinetic is therefore evaluated using the normalized area $a_g^{k+1}$ measured 
from the reconstructed field. 

The same 50 initial fields are simulated under two prescribed coefficients,
$M_{\mathrm{eff}}=5.0\times10^{-4}$ and $7.5\times10^{-4}$. Each simulation 
comprises 50 steps with $\Delta t=0.5$. For each step observation, the normalized area-change
rate is
\begin{equation}
y_g^k=\frac{a_g^{k+1}-a_g^k}{\Delta t}.
\end{equation}
Grains touching the domain boundary in $L^k$ or satisfying $n_g^k<3$ are
excluded. If an otherwise eligible grain disappears during the step, its
area is assigned $a_g^{k+1}=0$. 

Two fits are calculated for every simulation. A stepwise fit uses all
eligible grains at a fixed step, producing $\widehat\beta_{1,j}^{\,k}$ 
for simulation $j$ and step $k$. A pooled fit combines all eligible grain-step 
observations from the 50 steps of one simulation, producing $\widehat\beta_{1,j}^{\,\mathrm{pool}}$. 
Thus, the stepwise fits quantify temporal variation in the fitted coefficient,
whereas the pooled fit provides one simulation-level estimate.

Fit quality is summarized by the coefficient of determination
\begin{equation}
R^2
=1-\frac{
\displaystyle\sum_{(g,k)\in\mathcal D}
\left(y_g^k-\widehat y_g^k\right)^2
}{
\displaystyle\sum_{(g,k)\in\mathcal D}
\left(y_g^k-\overline y_{\mathcal D}\right)^2
},
\end{equation}
where $\mathcal D$ denotes the observations include in the corresponding
stepwise or pooled fit, $\widehat y_g^k$ is the fitted rate and
$\overline y_{\mathcal D}$ is the mean observed rate over $\mathcal D$.

Ensemble summaries treat the initial field as the sampling unit. For coefficient 
condition $c$, let $\widehat\beta_{1,cj}$ denote the fitted slope from simulation $j$,
where $j=1,\ldots,50$. The ensemble mean and sample standard deviation are
\begin{equation}
\overline{\beta}_{1,c}
=\frac{1}{50}\sum_{j=1}^{50}\widehat\beta_{1,cj},\qquad s_{\beta_1,c}
=\left[\frac{1}{49}
\sum_{j=1}^{50}\left(\widehat\beta_{1,cj}-\overline{\beta}_{1,c}\right)^2
\right]^{1/2}.
\end{equation}
The corresponding normal-approximation 95\% confidence interval is
\begin{equation}
\overline{\beta}_{1,c}
\pm
1.96\frac{s_{\beta_1,c}}{\sqrt{50}}.
\end{equation}
For the pooled comparison, $\widehat\beta_{1,cj}$ denotes the slope fitted
over all steps of simulation $j$. For the stepwise comparison, the same
calculation is applied separately at each step using
$\widehat\beta_{1,cj}^{\,k}$. All 50 simulations are retained in both
coefficient ensembles.

\subsection{Coarsening observables}

All observables are calculated from the reconstructed field $L^k$. We denote the grain index at pixel $x$ and step $k$ as $I^k(x):\Omega\rightarrow\mathbb{N}$. Let $\mathcal G^k$ denote the set of grain indices present in the field 
and $N^k=|\mathcal G^k|$ the grain count. The area of grain $g$ and the mean grain area are
\begin{equation}
A_g^k=\sum_{x\in\Omega}\mathbf{1}\!\left[I^k(x)=g\right],
\qquad
\langle A^k\rangle=\frac{1}{N^k}\sum_{g\in\mathcal G^k}A_g^k ,
\end{equation}
where area is measured in pixels.

Let $\mathcal P_4$ be the set of unordered, horizontally or vertically adjacent pixel pairs within
the domain. The total grain-boundary length is
\begin{equation}
L_{\mathrm{bnd}}^k=
\sum_{\{x,y\}\in\mathcal P_4}
\mathbf{1}\!\left[I^k(x)\neq I^k(y)\right],
\end{equation}
so each unlike pixel pair contributes one pixel unit of boundary length.

Curvature is evaluated only for analysis. For each grain $g$, a signed-distance field that is positive
inside the grain is smoothed with a Gaussian kernel of standard deviation $\sigma=1.5$ pixels,
giving the scalar field variable $\phi_g^k$. The curvature field is
\begin{equation}
\kappa_g^k=\nabla\!\cdot\!\left(\frac{\nabla\phi_g^k}{|\nabla\phi_g^k|+10^{-8}}\right).
\end{equation}
Grains occupying fewer than 20 pixels are excluded. If $\partial_4 g^k$ denotes the four-connected
boundary pixels of grain $g$, the reported mean absolute curvature is
\begin{equation}
\overline{|\kappa|}^{\,k}=
\frac{\displaystyle\sum_{g:A_g^k\geq4}\sum_{x\in\partial_4 g^k}|\kappa_g^k(x)|}
{\displaystyle\sum_{g:A_g^k\geq4}|\partial_4 g^k|}.
\end{equation}
This is a boundary-pixel-weighted mean rather than an equally weighted mean of the individual-grain
values. The smoothing and curvature calculation do not modify the field or influence subsequent
simulation steps.

The equivalent radius, scaled area and scaled radius are
\begin{equation}
r_g^k=\sqrt{\frac{A_g^k}{\pi}},\qquad
\alpha_g^k=\frac{A_g^k}{\langle A^k\rangle},\qquad
\rho_g^k=\frac{r_g^k}{\langle r^k\rangle},
\end{equation}
where $\langle r^k\rangle=N_k^{-1}\sum_{g\in\mathcal G^k}r_g^k$. The distributions of
$\alpha_g^k$, $\rho_g^k$ and $n_g^k$ are shown at steps 0, 20, 40 and 50 in the representative simulation. For the statistical
comparisons reported in the main text, two-sided two-sample Kolmogorov--Smirnov tests compare
scaled area and side number between successive displayed states: 0--20, 20--40 and 40--50.

\section{Phase-field benchmark protocol}
\label{app:benchmark}

\subsection{Phase field simulation setting}

The phase field benchmark is implemented using MOOSE. We compare computational times using five initial fields. 
The MOOSE \texttt{GBEvolution} material is parameterized by a grain-boundary energy of $0.708~\mathrm{J\,m^{-2}}$, 
a mobility prefactor of $5.0\times10^{-4}~\mathrm{m^4\,J^{-1}\,s^{-1}}$, an activation energy of $0.23~\mathrm{eV}$, 
a temperature of $450~\mathrm{K}$ and a diffuse grain-boundary width of 14 in the model length scale. These parameters 
are identical for all grains, corresponding to isotropic grain-boundary properties.

The computational domain is $[0,1000]\times[0,1000]$ and is discretized by a
$128\times128$ mesh of first-order four-node quadrilateral elements. Periodic boundary conditions are imposed in 
both coordinate directions. The coupled order-parameter equations are integrated using the second-order backward
differentiation formula (BDF2) and solved using a preconditioned Jacobian-free Newton--Krylov method with the 
BoomerAMG preconditioner. The initial time-step size is 0.1 and is adjusted using iteration-based adaptivity,
with growth and cutback factors of 1.5 and 0.5, respectively. The nonlinear absolute and relative tolerances are 
$10^{-9}$ and $10^{-6}$, and the linear relative tolerance is $10^{-4}$.

Phase-field and VN-GraphGG are paired according to similar grain counts; neither equal 
simulation time nor correspondence between individual grains is assumed. For the curvature 
comparison, both methods are analyzed using the same procedure: each binary grain indicator is
Gaussian filtered with $\sigma=1$ pixel, grains smaller than 20 pixels are
excluded, and curvature values are pooled over four-connected boundary pixels.
Filtering is used only for analysis and does not modify either simulation.

\subsection{Timing protocol}

Let $N^0$ be the grain count in the initial field and let the count threshold be $N^0/2$. 
For the VN-GraphGG, the wall time at this threshold is obtained by linear interpolation 
between the two recorded states whose grain counts bracket $N^0/2$. For the phase-field simulation,
the MOOSE CSV output is polled every 2~s and the run is terminated when the
\texttt{GrainTracker} count is first observed at or below $N^0/2$. The reported
phase-field wall time is therefore the elapsed time to this observed stopping
event and is not interpolated between output states.

Both methods are run on an AMD Ryzen Threadripper PRO 3955WX CPU using one
process and one CPU thread. For VN-GraphGG simulation, PyTorch
intra-operation and inter-operation thread counts and the OpenMP, MKL,
OpenBLAS and NumExpr thread counts are each set to one. Wall time includes
initialization, model or solver construction, evolution, diagnostic output and
stopping checks. The phase-field timing runs write CSV diagnostics only.
VN-GraphGG times are averaged over three repeats, whereas each phase-field
case is run once.

\section{Abnormal grain growth}
\label{app:agg}

For each initial field, the eligible set comprises grains that do not touch the
outer domain boundary at $k=0$. Therefore, at $f_{\mathrm{high}}=100\%$, all initially interior 
grains are selected, whereas initially boundary-touching grains retain the baseline coefficient.

For an AGG-positive simulation, let $k_{\mathrm{AGG}}$ be the earliest
above-threshold observation of any selected grain that subsequently satisfies
the sustained criterion. The normalized onset is
\begin{equation}
\frac{t_{\mathrm{AGG}}}{t_{50}}
=\frac{k_{\mathrm{AGG}}}{k_{50}}.
\end{equation}
The severity of a simulation is
\begin{equation}
\textit{max}\, R_A=\max_{\substack{0\leq k<k_{50}\\ i\in\mathcal I_{\mathrm{selected}}^k}} R_i^k,
\end{equation}
where $\mathcal I_{\mathrm{selected}}^k$ is the set of selected grains present
at step $k$. This maximum includes all selected grains, irrespective of whether
they satisfy the sustained criterion.

The pointwise 95\% bootstrap percentile intervals shown for the occurrence
fractions in Fig.~\ref{fig:agg-application} are calculated from 10,000
initial-field-clustered resamples. For each condition, we use three different seeds to sample the selected grains in
ten initial fields, and thus run 30 simulations. The three
selection-seed results are first averaged within each of the ten initial
fields. The ten field-level values are then sampled with replacement, and the
2.5th and 97.5th percentiles of the resampled means define the interval.

\end{document}